\documentclass{article}

    \PassOptionsToPackage{numbers, sort&compress}{natbib}

    \usepackage[final]{neurips_2023}

\usepackage[utf8]{inputenc} 
\usepackage[T1]{fontenc}    
\usepackage[hidelinks]{hyperref}       
\usepackage{url}            
\usepackage{booktabs}       
\usepackage{amsfonts}       
\usepackage{nicefrac}       
\usepackage{microtype}      
\usepackage{xcolor}         

\usepackage{graphicx}
\usepackage{caption}
\usepackage{subcaption}

\title{Data is Overrated: Perceptual Metrics Can Lead Learning in the Absence of Training Data}

\author{
  Tashi Namgyal$^{1}$ \quad Alexander Hepburn$^{1}$ \quad Raul Santos-Rodriguez$^{1}$\\
  \textbf{Valero Laparra}$^{2}$  \quad \textbf{Jesus Malo}$^{2}$
  \\
  $^1$University of Bristol \quad $^2$Universitat de Valencia\\
  \texttt{\{tashi.namgyal, alex.hepburn, enrsr\}@bristol.ac.uk}\\
  \texttt{\{valero.laparra, jesus.malo\}@uv.es}}

\begin{document}

\maketitle

\begin{abstract}  
Perceptual metrics are traditionally used to evaluate the quality of natural signals, such as images and audio. They are designed to mimic the perceptual behaviour of human observers and usually reflect structures found in natural signals. This motivates their use as loss functions for training generative models such that models will learn to capture the structure held in the metric. We take this idea to the extreme in the audio domain by training a compressive autoencoder to reconstruct uniform noise, in lieu of natural data. We show that training with perceptual losses improves the reconstruction of spectrograms and re-synthesized audio at test time over models trained with a standard Euclidean loss. This demonstrates better generalisation to unseen natural signals when using perceptual metrics.
\end{abstract}

\section{Introduction}

Deep generative models for vision and audio are becoming more widespread in production systems and so the process of reliably evaluating the quality of their outputs has become more important. The gold standard for measuring the quality of generative model outputs is to gather mean opinion scores from a group of human participants. However, this is a slow and expensive process that is not well suited to the fast iterative design process common in machine learning. Instead, we would like to replace this process with objective metrics that are well correlated to human quality ratings. Unfortunately, traditional training objectives such as mean squared error are not particularly well correlated to such ratings \cite{vinay2022evaluating}. Instead, metrics that better match human quality ratings can be designed, \emph{perceptual metrics}, by taking into account structural information common across natural signals and/or the structure of neurological pathways used in perception. For example, Structural Similarity (SSIM) \cite{wang2004image} captures structural information in natural images and the Normalized Laplacian Pyramid Distance (NLPD) \cite{laparra2016perceptual}  mimics characteristics of the visual pathway.

These are examples of a reference-based paradigm where a degraded signal is compared to a high quality reference, and their distance (or similarity) can be used to measure the perceived quality of the degraded version. 
Applying and tailoring these image quality metrics to spectrograms has recently been shown to have a better correlation with human ratings of audio quality than equivalent audio quality metrics \cite{namgyal2023what}.

Despite only being optimised and designed for aligning with human perceptual opinion, perceptual image metrics have been shown to capture statistical information about natural images~\cite{hepburn2022on,malo2010psychophysically}. This is in agreement with the \emph{efficient coding hypothesis}~\cite{barlow}, which suggests that our sensory system has adapted to minimise redundant information when processing stimuli. This can have an advantage when training machine learning models. For instance, perceptual metrics have been used as regularisation in image-to-image translation where a batch size of 1 leads to inaccurate gradient estimations, which the metrics protect against by including information about the distribution of natural images in the loss function~\citep{hepburn2020enforcing}. In order to evaluate if similar reasoning applies in the audio domain, we follow the same setup as ~\cite{hepburn2022on}. We compare the use of perceptual metrics versus a Euclidean metric to train compressive autoencoders, using uniform noise as input data. We show that models optimised with perceptual metrics are better than the Euclidean metric model at reconstructing natural audio, despite both being solely trained on uniform noise. 

\section{Methods}

In what follows we unpick the design choices behind our approach. We first briefly present each perceptual metric (SSIM and NLPD). We then discuss the autoencoder architecture that we use. We finally cover the training and test setup, where at training time we learn to reconstruct uniform noise, while at test time we use these models to reconstruct spectrograms and measure each model's reconstruction performance. Additionally, we train models on a music dataset for comparison.

\subsection{Perceptual Metrics}

While SSIM and NLPD were originally designed to evaluate differences between images, it has been shown that they can be used to evaluate differences in spectrograms obtaining good agreement with human ratings of audio quality \citep{namgyal2023what}. 

\paragraph{Structural Similarity} SSIM calculates several statistical features between local image patches, namely luminance, contrast and structure. The mean and variance of these features are compared between two images to give an overall similarity score. SSIM can be calculated across multiple scales, with the image resolution halved at each scale, to calculate the Multi-scale Structural Similarity (MS-SSIM) \cite{wang2003multiscale}, which increases the correlation with human quality ratings.

\paragraph{Normalized Laplacian Pyramid Distance} NLPD is a neurologically-inspired distance based on two processes found in the early stages of the visual and auditory pathways, namely linear filtering and local normalisation ~\cite{schwartz2001natural,laparra2010metric,willmore2023adaptation}. These processes reduce the redundancy found in natural signals in agreement with the efficient coding hypothesis \cite{malo2010psychophysically}. NLPD models these processes with a Laplacian Pyramid step \cite{burt1983laplacian} followed by a divisive normalization step \cite{malo2010psychophysically}. The resulting distance measure is well correlated with human perception of images \cite{laparra2016perceptual} and audio \cite{namgyal2023what}.

\subsection{Architecture} 

\begin{figure}[b]
    \centering
    \includegraphics[width=\textwidth]{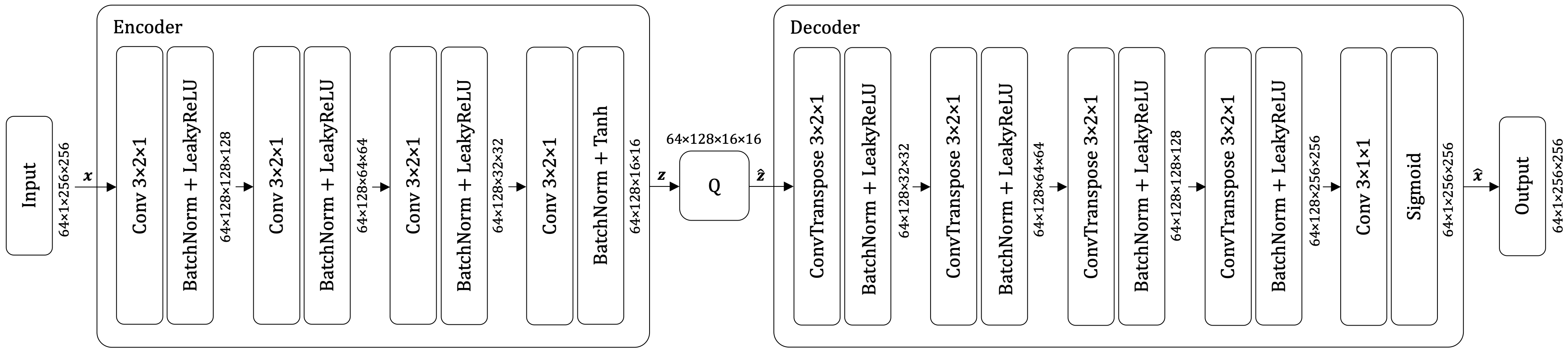}
    \caption{Autoencoder architecture. Dimensions are shown at each stage as batch size $\times$ channels $\times$ height $\times$ width. Convolution and batch normalization operations are 2-d. Convolution parameters are displayed as filter size $\times$ stride $\times$ padding. Q represents the quantisation step, shown in Eq. \ref{eq:quantisation}.}
    \label{fig:architecture}
\end{figure}

We use an autoencoder architecture with convolutional encoder and decoder networks and a quantised latent space. The encoder and decoder consist of four convolutional layers with 128 filter channels, which each scale the resolution by a factor of 2. Each convolutional layer is followed by batch normalisation and a LeakyReLU activation. In the final layer of the encoder, the LeakyReLU activation is replaced with a Tanh function to ensure the latent space is scaled to [-1,1]. The decoder has an additional final convolutional layer to reduce the number of channels from 128 to 1 and a Sigmoid activation to scale the output to [0,1]. The full architecture is shown in Fig. \ref{fig:architecture}. 

In order to control the level of compression, we constrain the embedding for each autoencoder to have a fixed maximum possible entropy. This ensures that the behaviour observed comes from the loss function used, rather than one embedding having higher entropy than another. To achieve this, we use a soft differentiable quantisation step that fixes the number of possible latent values~\citep{agustsson2019generative, ding2021comparison}:

\vspace{-0.5em}
\begin{equation}\label{eq:quantisation}
   \hat{z}_i = \sum_{j=1}^L \frac{ \exp \left( -s \left( z_i-c_j \right) ^2 \right) }{ \sum_{k=1}^L \exp \left( -s \left( z_i - c_k \right) ^2 \right)} c_j 
\end{equation}
\vspace{-0.5em}

\noindent where $c_1,...,c_L$ are the quantisation centres, $L$ is the number of centres, and $s$ is a scale factor. As $s$ is increased, the input values are drawn further towards the centres, but if $s$ is too large this may lead to numerical instability. We use $s=10$ and two centres $\{-1,1\}$. An upper bound on the entropy of the latent representation can then be calculated:

\vspace{-0.5em}
\begin{equation}
    \mathbb{H}\left( \hat{z}_i \right) \leq \frac{W \times H}{2^n \cdot 2^n} \cdot m \cdot \log_{2} \left( L \right)
\end{equation}
\vspace{-0.5em}

\noindent where $n$ is the number of layers of down-sampling, $W$ and $H$ are the width and height of the input, $m$ is the number of channels in the embedding and $L$ is the number of centres, as defined in Eq. \ref{eq:quantisation}. 
We use $n=4$, $W,H=256$, $m=128$ and $L=2$ for an upper bound on entropy in the embedding of $32768$ bits.
Further dividing by the input size gives the number of bits per pixel (bpp) required to encode the image. A standard RGB image is 24bpp (3 channels $\times$ 8 bit depth). The audio used here also has a bit depth of 24. We compress to 0.5bpp, which represents a compression ratio of 48:1.  

\subsection{Data}

\paragraph{Training} We train on two kinds of data, uniform noise in the range [0,1] and mel-spectrograms created from the MusicCaps\footnote{\url{https://research.google/resources/datasets/musiccaps}} dataset. The MusicCaps dataset consists of 10-second music clips with human generated captions, which we trim to the same length as the PMQD dataset and apply the same processing (see below). We use a batch size of 64, a single channel, $256 \times 256$ spatial dimensions and an AdamW optimiser with an initial learning rate of 0.001. Models are trained for 5 epochs (425 iterations) as models begin to over-fit beyond this. Training on spectrograms takes 18 minutes on a cluster of 4 A10G GPUs and training on noise takes 9 minutes on a single V100 GPU. 

\paragraph{Test} We use audio from the Perceived Music Quality Dataset (PMQD)\footnote{\url{https://github.com/carlthome/pmqd}} \cite{hilmkil2020perceiving}, which contains 195 4.208-second reference music clips. Audio clips are down-mixed from stereo to mono, down-sampled from 48kHz to 16kHz and converted to magnitude mel-spectrograms using a window size of 1024, hop length of 260 and 256 mel-bands, resulting in an output size of $256 \times 256$. Spectrograms are converted to a log-scale with an epsilon of $0.001$, to avoid undefined values, and scaled to lie in the range [0,1]. Audio is reconstructed from mel-spectrograms by reverse-scaling and passing through the Griffin-Lim algorithm. Transforms are performed with the TorchAudio package\footnote{\url{https://pytorch.org/audio/main/torchaudio}}.

\section{Results}

\begin{table}[h!]
  \caption{Comparison of reconstruction error according to three different metrics: MSE, NLPD and MS-SSIM. $P(x)$ indicates models trained on the MusicCaps dataset, $U(x)$ indicates models trained on uniform noise. The rows correspond to separate autoencoder models each trained with a different loss, as indicated in the `Loss' column. The columns under Test Metric show the distances/similarity on a batch of spectrograms from the PMQD dataset at test time. \vspace{0.5em}}
	\centering
	\begin{tabular}{|c|c|ccc|}
\hline
       &      & \multicolumn{3}{c|}{Test Metric}                                                                                                   \\ \cline{3-5} 
Training Data & Loss & \multicolumn{1}{c|}{MSE $\downarrow$}                  & \multicolumn{1}{c|}{NLPD $\downarrow$}               & MS-SSIM $\uparrow$                  \\ \hline
       & MSE  & \multicolumn{1}{c|}{\textbf{0.003119}} & \multicolumn{1}{c|}{3.738}                           & 0.9593                           \\
$P(x)$ & NLPD & \multicolumn{1}{c|}{0.008987}                           & \multicolumn{1}{c|}{\textbf{2.962}} & 0.9658                           \\
       & MS-SSIM & \multicolumn{1}{c|}{0.003528}                           & \multicolumn{1}{c|}{3.390}                           & \textbf{0.9717} \\ \hline
       & MSE  & \multicolumn{1}{c|}{0.03174}                            & \multicolumn{1}{c|}{7.504}                           & 0.8419                           \\
$U(x)$        & NLPD & \multicolumn{1}{c|}{\textbf{0.01124}} & \multicolumn{1}{c|}{\textbf{4.912}} & \textbf{0.8995} \\
       & MS-SSIM & \multicolumn{1}{c|}{0.02316}                            & \multicolumn{1}{c|}{5.999}                           & 0.8682                           \\ \hline
\end{tabular}
	\label{tab:losses}
\end{table}

Table \ref{tab:losses} shows the results of the models trained using each of the three metrics (MS-SSIM, NLPD, and MSE) on the noise and audio datasets.   
For the models trained using audio data, the greatest reduction in error corresponds with the metric used as the loss function. However, of the models trained on uniform noise and later presented with spectrograms, the models trained with NLPD and MS-SSIM have a lower MSE than the model trained with MSE as the loss function, demonstrating better generalisation to unseen natural signals when using perceptual metrics. 

Audio clips reconstructed with perceptual losses also sound more natural (audio examples for a variety of genres and musical styles are available\footnote{ \url{https://audio-perception.notion.site/Examples-4c3159f41b5e4c3eb8f2187fe5be55f5} }). The reconstructions are not expected to be perfect as the input mel-spectrograms are passed through a heavily compressed latent space. This makes the task of phase reconstruction particularly challenging, resulting in audible artifacts. However, these artifacts are less prominent in the models trained with perceptual metrics, especially NLPD. We find this to be the case for models trained on both noise and spectrograms, despite the disagreement with test metric scores on the spectrogram trained models. 

\begin{figure}[t]
    \centering
    \includegraphics[width=\textwidth]{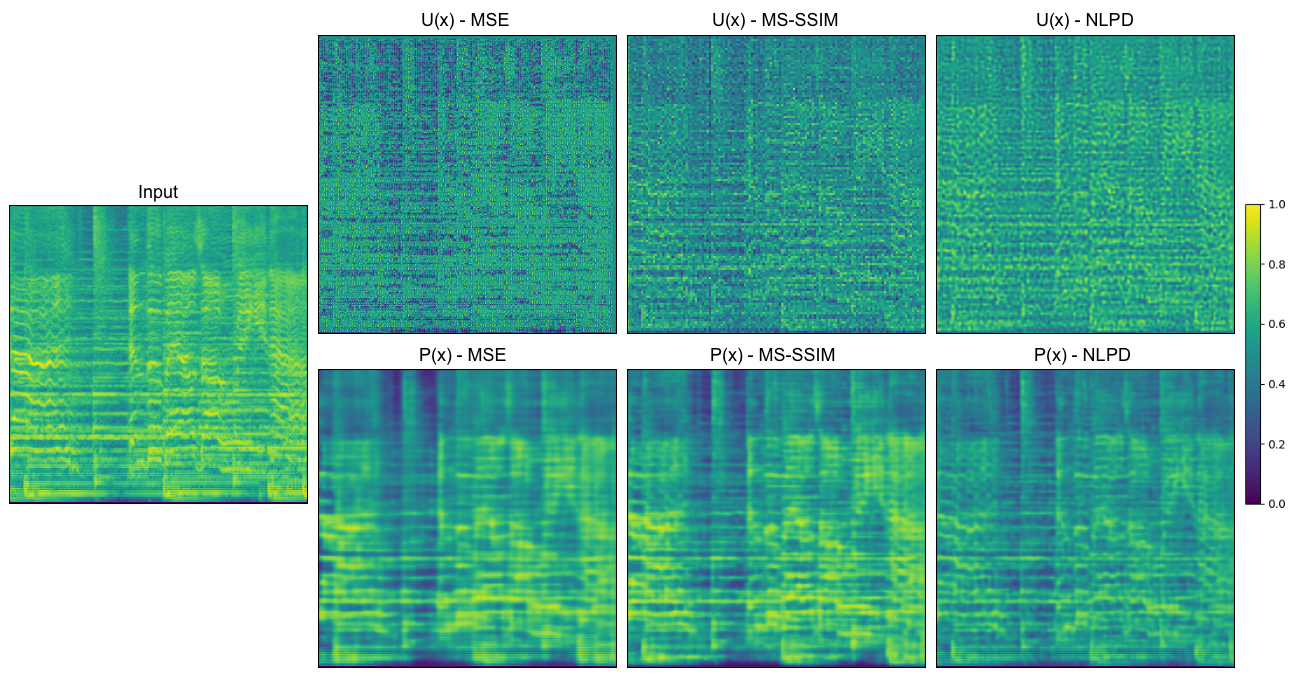}
    \caption{Input and reconstructed spectrograms from autoencoders trained with uniform noise $U(x)$ or music $P(x)$ using different loss functions: MSE, MS-SSIM and NLPD.}
    \label{fig:specs}
\end{figure}

Reconstructed spectrograms show different kinds of visual artifacts for each metric that match these audible artifacts (Fig. \ref{fig:specs}). For example, the uniform noise MSE model shows a flattening effect and has a harsh, noisy, ringing audio reconstruction. The spectrogram trained MSE model exhibits smoothing, which is particularly audible as smoothed percussive hits. The NLPD trained models reveal more detail, particularly for changing pitch and vibrato.

\section{Conclusion}
Perceptual metrics that are used across different modalities and domains are assumed to capture information about the structure of natural signals. As such, it is our assumption that using these metrics as loss functions when training machine learning models would help to better represent natural signals. In this work, we take this to the extreme by removing all structure in the input signal during the training phase, i.e. by training on uniform noise in lieu of natural data. Models trained in this way using a perceptual metric as a loss function have been shown to be able to reconstruct natural images at test time \cite{hepburn2022on}. Here we demonstrate that this is also the case for music. This paves the way for principled training and evaluation of generative audio models with perceptual metrics, while potentially alleviating the requirement for prohibitively large-scale datasets. 

\newpage

\section*{Acknowledgments}
TN is supported by the UKRI AI CDT (EP/S022937/1). AH and RSR are supported by UKRI Turing AI Fellowship EP/V024817/1. VL and JM are supported by MINCEO and ERDF grants PID2020-118071GB-I00, DPI2017-89867-C2-2-R and GV/2021/074.

\bibliographystyle{abbrvnat}
\bibliography{references}


\end{document}